Statistics → Applications → Epidemiology

*Does adjustment for measurement error induce positive bias if there is no true association?*

Igor Burstyn, Ph.D.

Community and Occupational Medicine Program, Department of Medicine, Faculty of Medicine and Dentistry, the University of Alberta, Edmonton, Alberta, Canada;

Tel: 780-492-3240; Fax: 780-492-9677; E-mail: iburstyn@ualberta.ca

**Abstract**

This article is a response to an off-the-record discussion that I had at an international meeting of epidemiologists. It centered on a concern, perhaps widely spread, that measurement error adjustment methods can induce positive bias in results of epidemiological studies when there is no true association. I trace the possible history of this supposition and test it in a simulation study of both continuous and binary health outcomes under a classical multiplicative measurement error model. A Bayesian measurement adjustment method is used. The main conclusion is that adjustment for the presumed measurement error does not 'induce' positive associations, especially if the focus of the interpretation of the result is taken away from the point estimate. This is in line with properties of earlier measurement error adjustment methods introduced to epidemiologists in the 1990's. An heuristic argument is provided to support the generalizability of this observation in the Bayesian framework. I find that when there is no true association, positive bias can only be induced by indefensible manipulation of the priors, such that they dominate the data. The misconception about bias induced by measurement error adjustment should be more clearly explained during the training of epidemiologists to ensure the appropriate (and wider) use of measurement error correction procedures. The simple message that can be derived from this paper is: 'Do not focus on point estimates, but mind the gap between boundaries that reflect variability in the estimate'. And of course: 'Treat measurement error as a tractable problem that deserves much more attention than just a qualitative (throw-away) discussion'.

**Introduction**

There is a suspicion among some epidemiologists that correction for error in exposure variables can artificially create an observed positive association. The typical argument to support this notion proceeds along these lines: "If non-differential exposure misclassification (measurement error) attenuates an estimate of risk, then adjusting for this phenomena will increase the risk estimate proportionately to the presumed extent of imprecision in the observed exposure. In most cases an argument can be made that the observed exposures were not influenced by the knowledge of health outcome and therefore, exposure misclassification (measurement error) is non-differential. By simply assuming an ever increasing magnitude of error in exposure, one can arrive at a correspondingly increasing risk estimate even if there is no true association." Such arguments are rarely voiced openly, but are a common concern in private discussions and are one reason put forward against the use of measurement error adjustment techniques. However, if "measurement error is threatening our profession"[1], then the apparent avoidance of measurement error adjustment techniques [2] seems to be a suicidal tendency for epidemiology. This article is an attempt to contribute to overcoming the reluctance of epidemiologists to explicitly tackle the measurement error problem.

The above perception of artificial inflation of exposure-response associations due to measurement error adjustment can perhaps be traced to some of the early measurement error correction methods introduced to epidemiologists, such as the following relationship, popularized by Armstrong [3], between the true ($\beta$) and observed ($\beta^*$; 'asterisk' denotes an observed parameter) slopes of linear regression via the coefficient of reliability ($\rho_{xx}$): $\beta = \beta^*/\rho_{xx}$, which assumes a non-differential classical additive measurement error model. A similar relationship is known for the true and the observed relative risks (RR): $RR \cong RR^{*(1/\rho_{xx})}$, under the assumptions stated above [4]. If we note that $\rho_{xx} < 1$ and is inversely proportional to the measurement error variance, then it is obvious that as the measurement error variance grows larger, the correction methods proposed above will yield ever increasing estimates of slopes and relative risks (association parameters). Consequently, a person wishing to 'game' the rules can always postulate a measurement error variance sufficiently large to arrive at some target 'elevated' association parameter. While there is no empirical evidence (to my knowledge) of abuse of the above equations, the obvious opportunity for a dishonest or naïve individual to bias the results seems to have cast a shadow of suspicion on all measurement error correction techniques.

The simple approaches to the measurement error problem reviewed above do not reflect the state of the art in measurement error adjustment. Some of the developments in the field are Bayesian methods that reconcile our knowledge about measurement error with the available data and do not simply 'fix' naïve estimates of slopes and relative risks by an external multiplier [5,6]. It must be noted that even the simple approaches to measurement error correction illustrated above do not alter significance testing and may in fact lead to the widening of confidence intervals without adding insights about the direction of the association [4,7]. Thus, the concern about falsifying findings with the correct application of the older approaches is also not justified, *if one properly considers the variability of the estimate in the interpretation*. In fact, it was shown in Bayesian framework "that failing to adjust for misclassification can (lead one to) overstate the evidence", but "an honest admission of uncertainty about the misclassification" can also result in a more accurate estimate of the association parameter; "neither of these phenomena are predicted by common rules-of-thumb"[8]. Overall, given the bias towards over-interpreting point estimates among epidemiologists [9,10], it is worth addressing the concern about positive bias in post-adjustment association parameter under the true null association.

The purpose of this article is to determine whether the slopes of linear and logistic regressions that are known to be flat could be biased upward by employing Bayesian correction for suspected multiplicative measurement error.

**Analysis**

I consider a prospective epidemiological study, a cohort, in which health outcomes are either binary or continuous and exposures, distributed log-normally are measured for each subject. The study, if properly analyzed, should reveal no association between the exposure and the outcomes.

*Statistical models*

I propose a measurement error model that seems to arise naturally in environmental and occupational epidemiology of chemical exposures and particles (and probably many other applications). True exposure (X) is assumed to be related to observed exposure (W) through a multiplicative error process: $W = X \times e$, such that $e \sim LN(0, \tau_e)$; i.e. a classical multiplicative measurement error model (NB: hereafter $\tau$. is the precision parameter equal to the inverse of variance). Since most environmental and occupational exposure cannot be less than zero, we also assume that $X \sim LN(\mu_x, \tau_x)$. Two health outcome models are examined. First, health outcome (Y) is assumed to be measured on a continuous scale that can be normalized: $Y \sim N(\mu_y, \tau_y)$. Second, a binary disease state (Z), given exposure X, is assumed to follow Bernoulli distribution: $Z \sim Bern(\pi)$; $\pi = p(Z=1|X)$. Independence is assumed among $\pi, \mu_y, \tau_y, \mu_x, \tau_x$ and $\tau_e$.

*Simulated data & naïve analyses*

I will focus on the large-sample performance of measurement error adjustment procedures, not on the variability in its performance due to random error in finite samples. This is consistent with the objective of investigating what is *expected* to happen to the association parameter when the null hypothesis is true. Consequently, a very large simulated cohort consisting of 100,000 subjects was generated only once. Each subject was assigned true exposure as $LN(0, 1)$ and observed exposures were simulated with $\tau_e = 1$. Continuous health outcomes were assigned to each subject by drawing samples from $N(\mu_y = 0, \tau_y = 1)$. A binary disease state was assigned to 5% of the cohort ($\pi = 0.05$). Naïve frequentist models were fitted to the resulting data to test whether observed exposure W is associated with either Y or Z, using simple linear and logistic regressions, respectively. If true exposure X was observed, these models would be: $Y|X = \beta_0 + \beta X + \varepsilon$, where $\varepsilon \sim N(0, \sigma_\varepsilon^2)$ and $logit(\pi)|X = \alpha_0 + \alpha X$; of course $E(\hat{\beta}) = E(\hat{\alpha}) = 0$: exposure and outcome are unrelated. With surrogate W instead of X, the models become $Y|W = \beta^*_0 + \beta^* W + \varepsilon^*$, where $\varepsilon \sim N(0, \tau^*_\varepsilon)$ and $logit(\pi)|W = \alpha^*_0 + \alpha^* W$. Simulations and naïve analyses were undertaken in *R* environment [11].

*Bayesian adjustment for measurement error*

To simplify the notation, let [•] denote the probability density function of random variable •, as well as define $\boldsymbol{\beta} = (\beta_0, \beta)$ and $\boldsymbol{\alpha} = (\alpha_0, \alpha)$. I estimated naïve models with the slope and the intercept for [Y|W] and [Z|W] using frequentist procedures. Next, I formulated Bayesian models that ignore measurement error (not described in detail) and those required to correct for measurement error in W. To accomplish the latter task, I specified three models [5,12]:

1. Outcome/disease model ($[Y|X, \boldsymbol{\beta}, \tau_\varepsilon]$ or $[Z|X, \boldsymbol{\alpha}]$),
2. Measurement error model $[W|X, \tau_e]$, and
3. Model for true exposure $[X|\mu_x \tau_x]$.

The above three models are specified using the description that appears in the previous sections and are linked by conditional independence assumptions. Using conditional independence assumptions, I derived the following proportionality for the posteriors:

$[\boldsymbol{\beta}, X, \tau_\varepsilon, \tau_e|W, Y] \propto [\boldsymbol{\beta}|Y, X, \tau_\varepsilon][W|X, \tau_e][\boldsymbol{\beta}][\tau_\varepsilon][\mu_x][\tau_x][\tau_e]$

and

$[\boldsymbol{\alpha}, X, \tau_e|W, Z] \propto [\boldsymbol{\alpha}|Z, X][W|X, \tau_e][\boldsymbol{\alpha}][\mu_x][\tau_x][\tau_e]$,

for the linear and logistic outcome models respectively.

The targets of inference are the conditional posterior distributions [β|rest] and [α|rest]. Samples from the posteriors were obtained by Markov chain Monte Carlo (MCMC) sampling in WinBUGS expert system, which selects appropriate estimation algorithms for the specified models and priors [13,14]. I chose the majority of priors to be uninformative. Priors had to be somewhat different for linear and logistic disease models for computational

reasons. For the linear disease model: $[\beta_0] \sim N(0, 100)$, $[\beta] \sim N(0, 100)$, $[\mu_x] \sim LN(0, 100)$, $[\tau_x] \sim \Gamma(0.01, 10)$, $[\tau_\varepsilon] \sim \Gamma(0.01, 10)$. For the logistic disease model: $[\alpha_0] \sim N(0, 10)$, $[\alpha] \sim N(0, 10)$, $[\mu_x] \sim N(0, 10)$, $[\tau_x] \sim \Gamma(0.1, 10)$. (All normally distributed priors are specified in N(mean, variance) notation in the previous two sentences). Next, I alternated between four types of prior distribution of the precision of measurement error, $[\tau_e]$ (Table 1). Smaller means and variances of priors on precision of measurement error reflect an ever increasing belief that large measurement error is present.

Three MCMC chains were obtained for each model using different initial values for $\beta$ and $\alpha$, while initializing other parameters at their expected (by simulation) values (NB: logistic regression proved to be much more sensitive to initial values in the sense that WinBUGS 'crashed' unless 'suitable' initial values were chosen). After 10,000 burn-in iterations, 40,000 samples from the posterior were used to summarize the posteriors in terms of means and 95% credible intervals (CrI). Convergence of MCMC chains was judged by visual inspection and by the potential scale reduction factor, $\hat{R} \approx 1$ at convergence (the square root of the ratio of the between-chain and the within-chain variances) [15]. WinBUGS was run from R environment [11] using R2WinBUGS function [16]. All computer codes required to reproduce the results are available from the author upon request (with the cautionary note on the very computer-intensive nature of the Bayesian models estimated for this paper when the dataset and number of MCMC iterations are both large).

*Parameter estimates in the simulated data*

The naïve frequentist estimate of change in Y per unit of W ($\hat{\beta}^*$) was 0.0001 with a 95% confidence interval from -0.0009 to 0.001. Without 'correction' for measurement error the posterior mean of slope ($\beta$) was 0.0001 with 95% CrI from -0.0008 to 0.001 (3,006 samples from the posterior). The naïve frequentist estimate of the odds ratio ($\exp(\hat{\alpha}^*)$) was 0.996 with a 95% confidence interval from 0.991 to 1.001 and the analogous descriptive statistic for the posterior of the odds ratio did not change when the Bayesian equivalent of the naïve model was fitted to the data. These results provide good evidence that the 'observed' exposure is not associated with the outcomes and demonstrate agreement of frequentist and Bayesian methods under uninformative priors.

After 'correction' for measurement error most of the point estimates of the slope in the linear disease model were <0.01 with credible intervals nearly centered on zero, except for that for prior type B: elevated posterior mean with a credible interval that includes zero, but noticeably shifted towards positive values (Table 2). In the logistic disease model, an ever increasing assertion about the presence of large measurement error appears to produce point estimates (to be more precise: posterior means) of odds ratios that are biased further and further away from the 'true value' of one: all credible intervals are very wide and include one (Table 3). In both the linear and logistic models, the posterior means of the association parameters are biased away from the estimates obtained in models that do not assume measurement error as would be expected if the measurement error correction procedure was 'nudging' the estimate away from the null. It is clear that over-interpreting the trend in point estimates in response to measurement error correction, in this case, would lead to a false positive conclusion. It must be noted that if the independence of exposure and outcomes was not known to be true by simulation in my examples, the lack of power would not be a plausible explanation for not excluding an association in these cases, since the simulated cohort is very large. Therefore, Bayesian adjustment for measurement error in the studied situations leads to the correct conclusion that true exposure does not appear to be associated with the two outcomes (or at least, for a cautious analyst, that measurement error is not the cause for the lack of the associations).

*Heuristic argument*

The conclusions that can be drawn from this simulation study apply, strictly speaking, to only those models and parameter estimates that were considered. Nonetheless, the results do suggest some general lessons about the likelihood that correctly applied measurement error adjustment methods may induce bias away from the null when there is no exposure-response association. A simplistic heuristic argument may help further reassure us that the conclusions drawn from the simulations can be generalized.

Let $U = f(b, V)$ be the model under consideration with respect to some (U, V) data with association parameter (b) in disease mode $f$, such that b>0 implies a positive association of U and V, while b=0 implies independence of U and V (which is indeed the case). A strong a priori belief that p(b>0) can be a function of a presumed non-differential measurement error model which is believed to attenuate the observed borderline positive association (b*), such that b>b*>0. A similar belief typically motivates measurement error adjustment: a suspicion that the 'true' association was missed and/or under-estimated due to errors in an exposure estimate. Suppose that we want to determine p(b>0|data) and p(b=0|data) and compare these probabilities to find whether there is support in the data for a positive association. Bayes theorem states that

p(b>0|data) = p(data|b>0)p(b>0) / p(data) (positive exposure-response)

and

p(b=0|data) = p(data|b=0)p(b=0) / p(data) (no exposure-response).

The comparison of interest is characterized by

$\Delta$ = p(data|b=0)p(b=0) / p(data|b>0)p(b>0),

i.e. is it more likely, given the data, that there is a positive association or that there is no association.

If U and V are independent, then p(data|b=0)>>p(data|b>0)$\cong$0, i.e. the likelihood that data arose under b=0 is much greater than the likelihood that data arose under b>0. In fact, if p(data|b>0) $\rightarrow$ 0+ (or =0 for the infinite sample size), then p(b>0|data) $\rightarrow$ 0+ (in a finite sample and tending to zero for the infinite sample size). In other words, no matter how strongly we express our belief that there is a positive association, for example because we suspect V to be subject to severe measurement error that attenuated b*, if the data strongly rejects the positive association, the calculations will favor the correct conclusion, i.e. $\Delta$>>1 and b=0. The only other way to bias $\Delta$ is to ensure that (p(b>0)/p(b=0)) > (p(data|b=0)/p(data|b>0)) by placing very little a priori faith into the null compared to the positive association. But if p(data|b>0) $\rightarrow$ 0+ then one would have to argue strongly in favor of p(b=0) $\rightarrow$ 0+ (null hypothesis all but rejected *a priori*). However, priors that do not allow for null association and/or completely dominate the data are a clear violation of the principles of empiricism and any such formulation should be noticed and rejected by the scientific community. In other words, a reasonably conducted Bayesian analysis will not induce an association that is not at all supported by the data under the presumed model.

**Conclusions**

The measurement error adjustment methods explored in this paper do not show any tendency to inflate an exposure-response association.  However, the same care should be taken in interpreting such adjusted association parameters as ought to go into considering the validity of inferences drawn from naïve estimates, with the added comfort of knowing that the impact of measurement error on the results has been reduced.

It is also worth re-iterating the cautionary note of Armstrong [4]: "If corrections are carried out on the basis of *incorrect* information on error magnitude, bias may be increased, rather than decreased."  The emphasis in the above quote in italics in the quote is mine as it reinforces the notion that only incorrect information about measurement error will induce bias.  In Bayesian methodology for measurement error, one is allowed to be uncertain about the extent of measurement error and the exact knowledge of the true distribution of exposure is not necessary[6].  I observed that the prior on measurement error variance did in fact influence the central tendency of the posterior distributions of the association parameters.  Therefore, while blatantly incorrect assumptions about measurement error structure and magnitude are likely to lead to biased inferences, Bayesian methods for measurement error adjustment appear to be able to reflect uncertainty about the magnitude of measurement error in the estimates that they yield, while still providing informative results.  In other words, in the Bayesian framework an investigator no longer has to rely on correct adjustment for getting one number right – the reliability coefficient for example – which is indeed a risky proposition.  It must be noted that even though the posterior distribution of the association parameters were influenced by the presumed extent of measurement error, the interpretation of the results that considers variability in the posterior sample consistently led to the correct overall conclusion of no association.

Of course in a small (finite) sample, spurious associations can arise which are absent in the population, but this is an independent phenomena from any (perceived) bias due to adjustment for measurement error.  Thus, any biases inherent in the unadjusted estimate will be present in the adjusted one unless care is taken to remove such bias, e.g. by simultaneous adjustment for latent confounding [17-20] and known (non-measurement-error) bias [21].  The simple and unoriginal [9] message that can be derived from this paper is: 'Do not focus on point estimates, but mind the gap between boundaries that reflect variability in the estimate'.  And of course: 'Treat measurement error as a tractable problem that deserves much more attention than just a qualitative (throw-away) discussion'.

**Conflicts of interest**

None

**Acknowledgements**

The author was supported by a Population Health Investigator salary award from the Alberta Heritage Foundation for Medical Research.  The author is thankful to Dr. David Lunn of Medical Research Council Biostatistics Unit, Institute of Public Health, Cambridge, UK, for assistance with implementation in WinBUGS.  Michele P. Hamm edited the manuscript.

**Table 1: Priors of the precision of measurement error model ($[\tau_e]$)**

|  | *Distribution*[1] | *Mean* | *Variance* |
|---|---|---|---|
| Uninformative | $\Gamma(0.1, 10)$ | 1 | 10 |
| Informative type A (close to true error value) | $\Gamma(1, 1)$ | 1 | 1 |
| Informative type B (inflated error) | $\Gamma(0.5, 1)$ | 0.5 | 0.5 |
| Informative type C (inflated error) | $\Gamma(0.05, 1)$ | 0.05 | 0.05 |

1: Gamma distributions in the form of ($k$ = shape, $\theta$ = scale) with mean = $k\theta$ and variance = $k\theta^2$.

**Table 2: Summaries of posteriors for linear outcome model under different priors of the precision of measurement error ($[\tau_e]$):** An ever greater measurement error is assumed with an increasing confidence as we move down the table from uninformative to informative type C prior; the sample size from converged MCMC chains used to calculate posterior mean and percentiles is 1,002.

|  | Summary statistics for posterior of slope ($\beta$) | | |
|---|---|---|---|
| $[\tau_e]$ | Mean | 2.5%ile | 97.5%ile |
| Uniformative[1] | 0.003 | -0.02 | 0.04 |
| Informative type A[2] | 0.002 | -0.03 | 0.03 |
| Informative type B[3] | 0.5 | -1.5 | 3.0 |
| Informative type C[4] | 0.0006 | -0.01 | 0.01 |

1: $\Gamma(0.1, 10)$; 2: $\Gamma(1, 1)$; 3: $\Gamma(0.5, 1)$; 4: $\Gamma(0.05, 1)$.

**Table 3: Summaries of posteriors for logistic outcome model under different priors of the precision of measurement error ($[\tau_e]$):** An ever greater measurement error is assumed with an increasing confidence as we move down the table from uninformative to informative type C prior; the sample size from converged MCMC chains used to calculate posterior mean and percentiles is 1,002

|  | Summary statistics for posterior of odds ratio ($\exp(\alpha)$) | | |
|---|---|---|---|
| $[\tau_e]$ | Mean | 2.5%ile | 97.5%ile |
| Uniformative[1] | 0.705 | 0.148 | 4.116 |
| Informative type A[2] | 0.780 | 0.233 | 2.312 |
| Informative type B[3] | 0.645 | 0.113 | 1.391 |
| Informative type C[4] | 0.559 | 0.082 | 1.822 |

1: $\Gamma(0.1, 10)$; 2: $\Gamma(1, 1)$; 3: $\Gamma(0.5, 1)$; 4: $\Gamma(0.05, 1)$.